\documentclass[11pt]{article}
\usepackage{float}
\usepackage{amsmath}
\usepackage{amssymb}
\usepackage{subfigure}
\usepackage{algorithm}
\usepackage[noend]{algorithmic}
\usepackage{graphicx}

\usepackage{epstopdf}

\usepackage{cite}
\usepackage{color} 

\usepackage[labelfont=bf,labelsep=period,justification=raggedright]{caption}

\makeatletter
\renewcommand{\@biblabel}[1]{\quad#1.}
\makeatother

\date{}

\begin{document}

\begin{flushleft}
{\Large
\textbf{Influence of media on collective debates}
}
\\
Walter Quattrociocchi$^{1,2,3}$, 
Guido Caldarelli$^{3,2,4}$,
Antonio Scala$^{4,2,3\,\,\ast}$
\\
\bf{1} Laboratory for the modeling of biological and socio-technical systems, Northeastern University, Boston, MA 02115 USA
\\
\bf{2} LIMS the London Institute of Mathematical Sciences, 22 South Audley St 
Mayfair London W1K 2NY, UK
\\
\bf{3} IMT Alti Studi Lucca, piazza S. Ponziano 6, 55100 Lucca, Italy
\\
\bf{4} ISC-CNR Physics Dept., Univ. "La Sapienza" Piazzale Moro 5, 00185 Roma, Italy
\\
$\ast$ E-mail: Corresponding antonio.scala@cnr.it
\end{flushleft}


\section*{Abstract}
The information system (T.V., newspapers, blogs, social network platforms) and its inner dynamics play a fundamental role on the evolution of collective debates and thus on the public opinion. 
In this work we address such a process focusing on how the current inner strategies of the information system (competition, customer satisfaction) once combined with the gossip may affect the opinions' dynamics.
A reinforcement effect is particularly evident in the social network platforms where several and incompatible "cultures" coexist (e.g, pro or against the existence of chemical trails and reptilians, the new world order conspiracy and so forth). 
We introduce a computational model of opinion dynamics which accounts for the coexistence of media and gossip as separated but interdependent mechanisms influencing the opinions' evolution.  Individuals may change their opinions under the contemporary pressure of the information supplied by the media and the opinions of their social contacts.
We stress the effect of the media communication patterns by considering both the simple case where each medium mimics the behavior of the most successful one (in order to maximize the audience) and the case where there is polarization and thus competition among media reported information (in order to preserve and satisfy their segmented audience). Finally, we first model the information cycle as in the case of traditional main stream media -- i.e, when every medium knows about the format of all the others -- and then, to account for the effect of the Internet, on more complex connectivity patterns -- as in the case of the web based information.
We show that multiple and polarized information sources lead to stable configurations where several and distant opinions coexist.

\section*{Introduction}

Nowadays, there is an ongoing intense scientific debate around the definition of the foundational concepts as well as about the most appropriate methodological approaches to deal with the understanding of social dynamics \cite{Lazer09,conte2012manifesto}.
However, the challenge of understanding human behavior remains complex and intricate. 
Humans are intentional (and not necessarily rational) and the their dynamics in the social space are influenced by the surrounding social context and even by the information reported on the media. 
Tv, newspapers, blogs act on the \textit{memetic} diffusion which, in turn, is affected by a massive amount of individual and social factors -- e.g. tastes, desires, goals, trust, social pressure etc.
To understand such mechanisms we have to consider several factors from how information is produced, up to how the various information sources interact within them (and with respect to the audience). 
Since such an interaction manifests in terms of production and selection of proposed contents, we introduce a novel model of opinion dynamics on coupled and interacting networks. We stress the role of the media content production strategies by considering both the simple case where each media mimics the behavior of the most successful one (in order to maximize the audience) and the case where there is polarization and thus competition among media reported information (in order to preserve and satisfy the segmented audience and to follow the editorial line).

The increasingly massive use of Internet as a source of information and as a medium of communication has lead to a shift of paradigm in the production/diffusion of contents as well as in the communication process. 
The debate about social relevant issues spreads and persists over the web by leading to the emergence of unprecedented social phenomena such as the massive recruitment of people around common interests, ideas or political visions \cite{Guillory2011,Bekkers2011,Moreno2011,Garcia2012}.
In the past years an intensive research effort has been payed in understanding social phenomena from innovation diffusion, to social influence, up to opinions and their dynamics \cite{Castellano2007,Lorenz2007,eRep,Latane90,Masonetal.,lavineLatane,festinger50,Redner2011,Maxwell,brunetti2011dynamic,brunetti2012minimum}; some of them have focused on the role of media \cite{QuattrociocchiCL11,Phillips83,knut02,Martins2010,Carletti2006,QuattrociocchiPC09} and of the web \cite{ugander2012,BorgeHolthoefer:2011p5170,ungerleider11-1}.

The interaction among media, with the advent of the WWW, has been subject to an important change: people are not passive anymore, but can be proactive to an extent that often main stream media acquire information directly by common people.
Main stream media compete for the audience and therefore interact by adjusting their format/contents to collect the highest number of followers. 
Hence, if on the one hand people get informed by the media, on the other hand the information (as well as the way they are reported) are even more influenced by the evolution of the mass tastes.
Media respond to the their editors which often are politically lined up and then we can have the emergence of monopoles (as in the case of regimes) or to {\em polarized groups} (as in the modern democracies) of information broadcasters aiming at influencing people toward one or another political party \cite{Deighton95,QuattrociocchiCL11}.
The aim of this work is a) to introduce in the field of opinion dynamics the role of the media dynamics as a results of a competition/imitation process which has the goal to reach the highest number of followers; b) to highlight the changes induced by the historical evolution of the information system from the traditional main stream media to the WWW and c) to study the effect of aggregation/fragmentation of opinions in mixed communication environment.

Since we want to stress the role of trust with respect to information available to an individual, we assume that the gossipers interact with their neighbors and with the media using the {\em bounded confidence model} (BCM) \cite{amblard01} -- i.e, only if the distance between their internal state (opinion) and the received information is below a given threshold $\sigma$ (tolerance) they will be more likely to adjust their own opinion. The higher the tolerance, the more the people are likely to be influenced by (because they trust) the information circulating.
On the other hand, the media aim to reach the highest number of followers, hence they change their message according to the attitude of the media with the highest number of followers.
Finally, we introduce competition among media memes \cite{Adamic05thepolitical,weng:competition} -- i.e., we mark the edges between media with positive and negative values causing respectively to converge or diverge respect to their neighbor's attitudes. We show that multiple and polarized information sources can lead to stationary configurations where several opinions coexist.

\section*{The model}
\begin{figure}[H]
 \centering
     {\includegraphics[width=.25\textwidth]{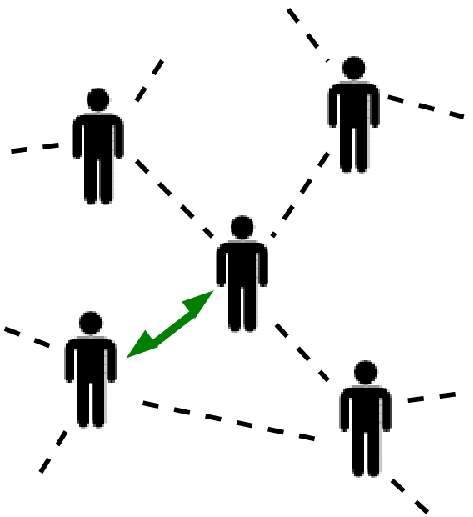} \label{subfig:gg}} 
   {\includegraphics[width=.25\textwidth]{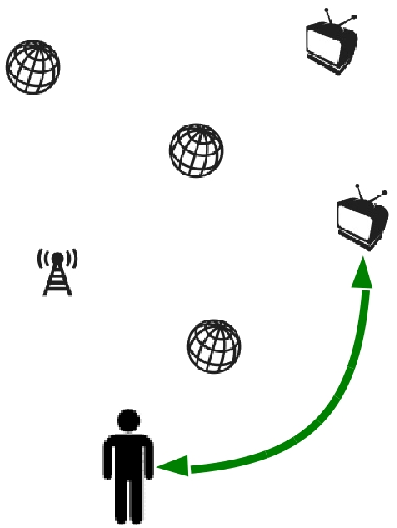} \label{subfig:gm}} 
   {\includegraphics[width=.30\textwidth]{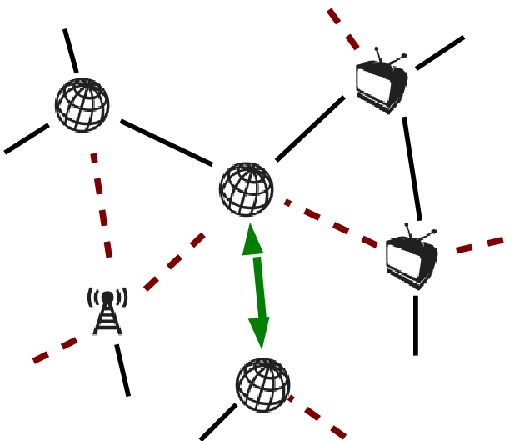} \label{subfig:mm}} 
   \caption{
A graphical sketch of our model. 
(Left panel) Gossipers interact among themselves choosing a neighbor in their social network (double arrow). If the gossipers have similar ideas, their opinion will converge further (eq.\protect{\ref{eq:gossipgossip}}).
(Central panel) Gossipers are also influenced by the media: when they are exposed to information, their opinion will converge to such information if it is not too far from the gossiper's initial opinion (eq.\protect{\ref{eq:gossipmedia}}).
(Right panel) Media are subject to a leader-follower dynamics. Media are supposed to have a network of other media with which interact either trying to copy their memes (black lines) or trying to oppose their memes (red dashed lines). Each media chooses to mimic/oppose the most successful (the one with more followers) of its neighboring media (eq.\protect{\ref{eq:mediagossip}}).
   }
\label{fig:model}
\end{figure}

The model discussed in this paper relies on two interacting networks: the media, which have the goal to collect the highest number of followers, and the gossipers, which can acquire information both from other gossipers and by the media. For the sake of simplicity, we imagine that the frequencies with which gossipers exchange opinions among themselves and consult the media are the same; thus, at each time-step both gossipers and media can adjust their opinions (figure \ref{fig:model}).
We assume that gossipers are likely to adjust their opinion only if the received information and their own beliefs are close enough, a situation referred as bounded confidence. 
Media, assumed to be more audience oriented, try to mimic the most successfull medium, i.e. the ones with the largest number of followers. The number of followers, i.e. of individuals which accept the information reported on by the medium, is updated at each time step. 
Every gossipers and media are initially assigned a random opinion described by a real value within a given opinion space $[0..1]$. 
Mathematical models of opinion dynamics under bounded confidence have been introduced by Axelrod in \cite{axelrod97} and then developed by Deffuant and Weisbuch \cite{amblard01} and by Hegselmann and Krause \cite{Hegselmann02opiniondynamics}.

We consider two interacting networks: the gossiper network $G_g$ and the media network $G_m$. Gossipers interact through the bounded confidence model (BCM) of Deffuant and Weisbuch \cite{amblard01} - i.e., at each time step $t$ a gossiper $i$ chooses at random a neighbor $j$ in its social network and adjusts its opinion according to

\begin{equation}
\label{eq:gossipgossip}
x_{i}^{t+1}=x_{i}^{t}+\mu_{gg}\left(x_{j}^{t}-x_{i}^{t}\right)\theta\left(\sigma_{gg}-\vert x_{j}^{t}-x_{i}^{t}\vert\right)
\end{equation}

where $x_{i}$ is the opinion of the gossipers $i$, $\mu_{gg}$ is a convergence factor,  $\sigma_{gg}$ is the threshold (opinion distance) above which gossipers do not interact and $\theta(\,\,)$ is Heaviside's theta function.

We assume that also the interaction with the media has a BCM form: 

\begin{equation}
\label{eq:gossipmedia}
x_{i}^{t+1}=x_{i}^{t}+\mu_{gm}\left(y_{k}^{t}-x_{i}^{t}\right)\theta\left(\sigma_{gm}-\vert y_{k}^{t}-x_{i}^{t}\vert\right)
\end{equation}

Here $k$ is a randomly chosen media, $y_{k}$ is the information (meme) reported by the
$k$-th media, $\mu_{gm}$ is a convergence factor and $\sigma_{gm}$ is the threshold below which gossipers gets influenced by the media. 

The media choice at time $t$ is described by the matrix $\xi_{ik}^{t}$ that is equals to $1$ if the $i$-th gossiper has chosen the $k$-th media, 0 otherwise; i.e. $\xi_{ik}^{t}$ is a binary random variable that takes the value $\xi_{ik}^{t}=1$ with probability $1/M$ and $\xi_{ik}^{t}=0$ otherwise. We can therefore count the followers of each media as

\begin{equation}
\label{eq:mediagossip}
f_{k}^{t}={\displaystyle \sum_{i}}\xi_{ik}^{t}\theta\left(\sigma_{gm}-\vert y_{k}^{t}-x_{i}^{t}\vert\right)
\end{equation}

where $\xi^{t}$ is calculated at each time-step.

We then introduce our max-audience oriented bounded confidence model among media interacting on a (possibly signed) network described by the matrix with elements $J_{kq}\in\left\{ -1,0,+1\right\}$. 
While the matrix $\vert J_{kq} \vert$ correspond to the adjacency matrix of the network, the sign of $J_{kq}$ indicates the polarization (friend/enemy) between the $k$-th and the $q$-the media. The case $J_{kq}\geq 0$ corresponds to unpolarized media.

First, the meme of the $k$-th media is influenced by the most successful (the leader) $l\left(k\right)$ of its neighbors  
$neigh(k)=\left\lbrace q : \vert J_{kq} \vert>0 \right\rbrace$
\begin{equation}
l\left(k\right)={\displaystyle \max_{q\in neigh(k)}}\left\{ \vert J_{kq}\vert f_{q}\right\} 
\end{equation}
and then its meme is updated accorded a signed version of the BCM model:
\begin{equation}
\label{eq:mediamedia}
y_{k}^{t+1}=\mathcal{B}\left[y_{k}^{t}+\mu_{mm}J_{kl\left(k\right)}\left(y_{l\left(k\right)}^{t}-y_{k}^{t}\right)\theta\left(\sigma_{mm}-\vert y_{l\left(k\right)}^{t}-y_{k}^{t}\vert\right)\right]
\end{equation}
where the function 
\begin{equation}
\mathcal{B}\left(y\right)=
\left\{ \begin{array}{c}
0\\
y\\
1
\end{array}\right.
\,\,if\,\,\,
\begin{array}{l}
y<0\\
0<y<1\\
y>1
\end{array}
\end{equation}

constrains the memes in the interval $\left[0,1\right]$; this is necessary as for $J_{kq}<0$ the memes among the $k$-th and the $q$-th media tend to diverge and could therefore go below $0$ or beyond $1$.

Notice that the convergence factors $\mu_{\alpha \beta}$ with $\alpha,\beta \in \left\{ g,m \right\}$
correspond to the timescales of the dynamics. 
In our study, we always use $\mu_{\alpha \beta}=0.3$ and $\sigma_{gg}=\sigma_{gm}=\sigma_{mg}=\sigma$

Since the opinion space $\left[0,1\right]$ is continuous, we can have different configuration in the final stationary opinion state. 
Opinions' clusters could be one (consensus), two (polarization), or more (fragmentation). 
In the following, we will first consider media as audience oriented agencies without any particular competition among them -- i.e., a situation in which the opinion of a medium converges to the most successful among its' neighbors. 
Then, we will introduce competition among media memes -- i.e., according to the structural balance of Heider \cite{heider}, we mark the edges between media with positive and negative values as in \cite{kleinberg2010,Wasserman1994,Leskovec2010} causing respectively a step toward or far from the most followed neighboring medium.

\section*{Results and Discussion}

An analysis reveals that the our model is not amenable of a simple analytical solution not even at the mean-field level (see Supporting information); therefore, we have resort on numerical investigations of the model.
The key for a correct understanding of how the dynamics works in our model is to evaluate the interplay of the mechanism of imitation at the local and global scales of the network. To shed light to the inner mechanisms of the dynamics, we will use very simple networks where the two levels are clearly discernible.

We performed a thorough simulation program which considers different connectivity patterns for the two interacting layers (gossipers and media networks). 
We first show the opinion dynamics in the case where gossipers are subject to the audience oriented media broadcasting. Then, we introduce competition (polarization) in the media dynamics - i.e, each node of the media network, depending on the edge signature (positive or negative), can diverge (or converge) to (or from) the value of the most followed media.
Both the audience-oriented and the competing media cases are first considered in the complete graph case. This is the case of traditional main stream media where everybody knows about everybody. 
Finally, we consider the model's dynamics when nodes of the media networks are linked through more complex connectivity patterns like in the case of the WWW.
 
The nodes of the gossipers network interact over a scale-free network generated with the Barabasi-Albert \cite{BarabasiAlbert} model that sets the exponent of the power law distribution to be $3.0$. We explored also other connectivity patterns (Watts-Strogatz small world networks \cite{watts98} with different rewiring probability) noticing that the underlying topology does not affect the model qualitative behavior.

We explore the model dynamics in terms of opinion distance and number of clusters as a function of the tolerance parameter. Each point of the parameter space (5000 steps) is averaged over 100 different possible initial configurations to attain the desired accuracy. When not specified, the size of the networks is to be assumed to be $10^4$ nodes.

\subsection*{Traditional main-stream media}

The size of actors in traditional main-stream media ($TMSM$ )is small: the number of televisions, radio stations, newspapers etc. allows everybody to check what the others are doing. 
For this reason, we will model interactions among the $TMSM$ as a complete graph.

We first analyze the trend of opinions' extremal distances $d$ -- i.e., the distance between the highest and the lowest opinions in the gossipers' network -- as a function of the tolerance $\sigma$. 
When $d=0$, all the gossipers have reached consensus and share the same opinion while for $d>0$, opinions are distributed; in the case of the standard BCM model, opinions are distributed in clusters (delta functions) separated by a distance higher than $\sigma$.

\begin{figure}[H]
\centering
\includegraphics[width=.65\textwidth]{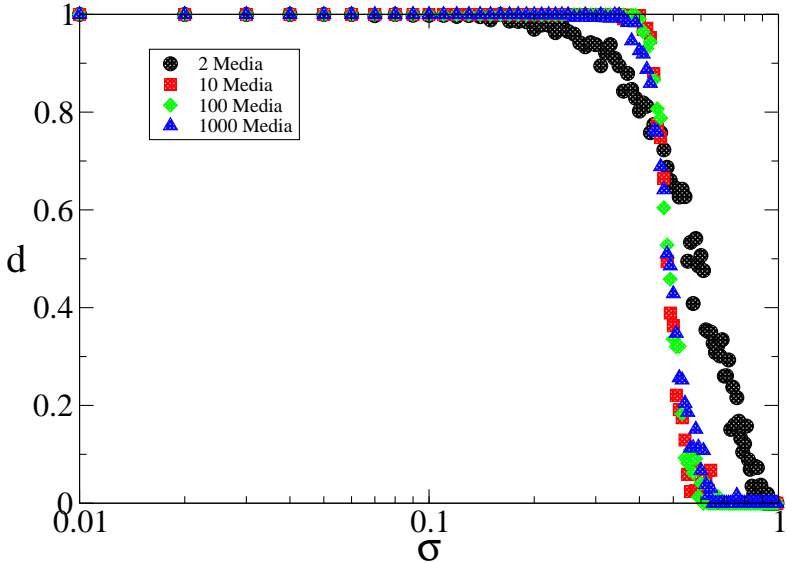}
\caption{Maximal opinion distance $d$ (the difference between the highest and lowest opinion in the gossipers' network) versus tolerance $\sigma$. The size of the simbols is bigger than the error bars. Opinions' distance under the effect of audience-oriented unpolarized media shows a smoothening of the transition (less sharp change of $d$ versus $\sigma$).  }
\label{fig:opdist_normal1}
\end{figure}

In Figure \ref{fig:opdist_normal1} we show the trend of the opinion distance (the highest opinion distance) as a function of the tolerance when the gossipers' network is scale-free \cite{BarabasiAlbert}. We have checked that qualitatively similar results hold for the other gossip network topologies. 
It is known that the increasing of the tolerance parameter causes a reduction of the distance within opinions until it reduces to {\em consensus} -- i.e, when the distance is $0$ -- and that the critical point (without media) is reached for value of tolerance of $0.5$ \cite{Fortunato2004a}.

Here, we show that the media action smoothens the transition. 
In fact, the presence of the media as audience oriented information agencies can enlarge the transition area before the consensus to a single opinion. Such an effect is more evident for a small number of media.
However, whether the interaction among different networks could change the order of the transition as in the case infrastructural networks \cite{BPPSH10} remains an open question.

The situation is completely different when we introduce competition (polarization) in the media information targeting mechanisms. 
In fact, in this scenario media can have negative or positive feedbacks from the other media; therefore the most popular medium can cause either a convergence (positive coupling) or divergence (negative coupling) of its neighbors' memes. 
For the sake of simplicity we set to $0.5$ the fraction of negative links of the media network -- i.e, the number of link that will cause divergence with respect to the most followed media message.

To measure the possible sparsity of opinions induced by the negative links, we bin the opinion space in a probability vector $\phi$ and measure the localization parameter \cite{NagelPRL1984}:
\begin{equation}
L = \frac{\left({\displaystyle \sum_{i}}\phi_{i}^{2}\right)^{2}}{{\displaystyle \sum_{i}\phi_{i}^{4}}}
\end{equation}
If $M_B$ is the (large) number of bins of the probability vector, $L\sim 1/M_B \sim 0$ if the opinions are evenly distributed, while $L=1$ if all the opinions are concentrated in a single bin. In general, if opinions are evenly distributed in $N_C$ clusters, $L^{-1}\sim N_C$ is of the order of the number of different opinions.

\begin{figure}[H]
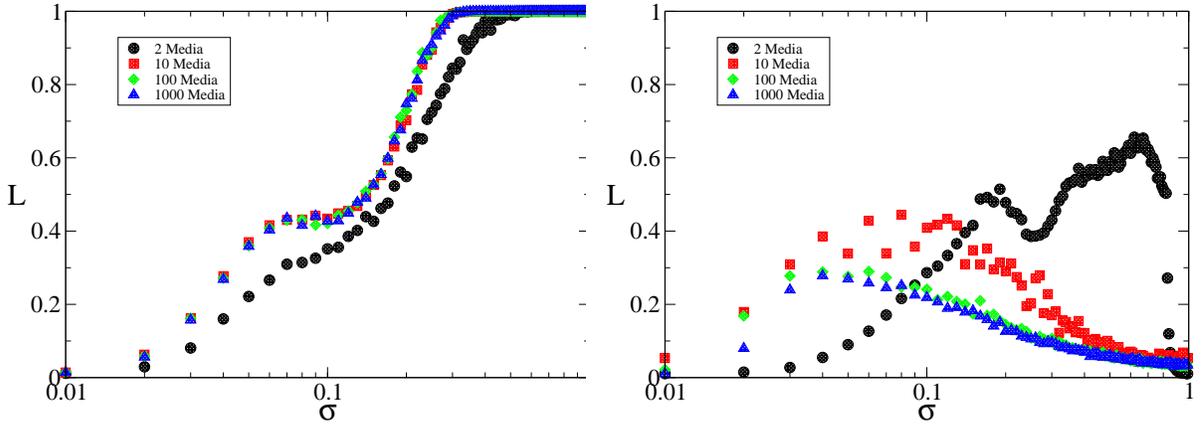

\centering
{\includegraphics[width=.49\textwidth]{3a}} 
{\includegraphics[width= .49\textwidth]{3b}}
\caption{Localization $L$ versus tolerance $\sigma$. Error bars are of the order of the symbol size.
The localization parameter can be thought as the inverse of the number of different opinion; therefore $L=1$ for consensus, while a low value of $L$ signals fragmentation of the opinions. 
(Left panel) Localization in gossip scale-free networks increases with the tolerance when the media are audience oriented agencies (i.e. unpolarized). Notice that full consensus ($L=1$) is reached at a tolerance $\sim 0.5$ like the single-network BCM model case.
(Right panel) Localization has a non-monotonic trend when media are polarized; in particular, it reaches a maximum \emph{before} the tolerance is maximal ($\sigma=1$). Like in the $BCM$ model, opinions are fragmented at low values of $\sigma$ since they do not interact; unlike the $BCM$ model, consensus is not reached at high values of $\sigma$ and opinions are fragmented due to the polarization of the media.}
\label{fig:opeaks}
\end{figure}

The left panel of Figure \ref{fig:opeaks} shows that for unsigned interactions among media, all the opinions converge to a single one, i.e. $L=1$ for high values of $\sigma$. 
Notice that in the case of a large number of unpolarised media, two opinion clusters ($L\sim 0.5$) cohexists at small $\sigma$'s (rougly between $\sigma=0.05$ and $\sigma=0.10$). 
The right panel of Figure \ref{fig:opeaks} shows a re-entrant effect in which the opinion space is maximally fragmented for both low and high values of the tolerance $\sigma$. However, such fragmentation stems from different mechanisms; in fact, while at low $\sigma$ opinion fragment into distinct non-interacting peaks as in the standard $BCM$, at values of tolerance the opinions fragment because the competition among polarised memes. In fact, at high $\sigma$ we find that while the average number of opinion peaks is stationary, their positions in the opinion space are dynamically fluctuating. Therefore, in the case of polarized media, even total trust in the information ($\sigma = 1$) does not produce full consensus as in the standard $BCM$, but induces a dynamically evolving stationary state in which actors can still change their opinions. Polarization triggers a sort of never-ending collective debate supported by different  and incompatible argumentations, not allowing people to find an agreement such as in the case of the existence or not of chemical trails or the link between vaccines and autism \cite{Kuklinski2000,Garrett2013}.

\subsection*{New media information}

\begin{figure}[H]
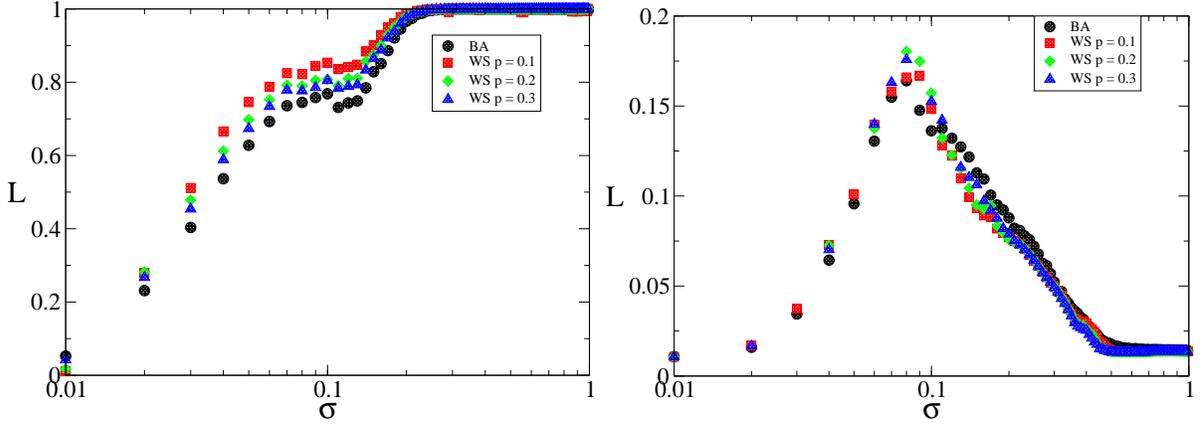

\centering
{\includegraphics[width= .49\textwidth]{4a}}
{\includegraphics[width= .49\textwidth]{4b}}
\caption{Localization $L$ versus tolerance $\sigma$ for Barabasi-Albert scale-free (BA) and Watts-Strogatz small world (WS) networks with rewiring probabilities $p=0.1,0.2,0.3$. Error bars are of the order of the symbol size.
The complex networks of gossipers and media are of comparable sizes. 
(Left panel) For un-polarized media, increasing the tolerance leads toward consensun ($L=1$). 
(Right panel) Polarization among the media produces a reentrant effect on the localization: while at low values of $\sigma$ opinions are fragmented since they do not interact, at high values of the tolerance polarization induces fragmentation. }
\label{fig:op_loc_complex} 
\end{figure}

The use of the Internet as a medium for sharing information has caused a shifting of paradigm from main stream centralized media to a more distributed and proactive protocol of information diffusion. At difference with traditional main-stream media, new media are composed of a large number of actors that cannot possibly check all the other media but that interact among each other through a social network.
Hence, in this last scenario we use different topologies for the media network -- i.e, of $10^4$ nodes with scale-free and small-world topologies (the latter with rewiring probabilities $p = 0.1, 0.2,0.3$)  -- acting on a gossip scale-free gossip network of the same size.
We analyze both polarized (audience oriented) and unpolarized (competing) media; in the former case, for the sake of simplicity we set to $0.5$ the fraction of negative links in the media network.
Figure \ref{fig:op_loc_complex} shows the behavior of the localization for scale free media networks and small world topologies. The high number of media makes extreme the effects evinced in the previous scenarios. 

We observe that the result for new media are qualitatively similar to the ones for the traditional main-stream media: in both cases competition among the media (polarization) introduces a non monotonic behavior in opinions' fragmentation, gossipers never reach consensus and the minimum opinion spread is reached for values of tolerance $\sigma <1$.

\section*{Conclusions}

In this paper we introduce a novel model of opinion dynamics accounting for the coexistence of media and of social influence as two separated but interdependent processes. People (nodes of the gossip network) interact with their neighbors or with the media using the Bounded Confidence Model (BCM) \cite{amblard01} -- i.e, only they will influence each other only if the distance between their opinions is below a given threshold $\sigma$ (tolerance), the higher $\sigma$, the more people are likely to be influenced.
In turn, the media (nodes of the media network) aim to capture the highest number of followers, hence they change their message by moving toward the value of the media with the highest number of followers. Finally, we introduce competition among media through polarization - i.e., media can interact with positive sign (their memes will converge) or negative sign (their memes will diverge).

We show that, when the media follow an audience oriented strategy of information delivery (i.e. everybody tries to mimic the most successful medium), there is a smoothening of consensus transition, indicating that the media messages tend to produce an impasse when amplified by the gossip dynamics. Such effect tends to disappear with increasing the number of media.
On the other hand, competition (polarization) among media produces a fragmentation of the opinions' space thus preventing a system-wide consensus. Such scenario is qualitatively robust to changes in the topologies of the gossip-gossip and medium-medium interaction network; in particular, it stays true both for conventional media (where everybody can interact with everybody) and for new media (where interactions take the form of a social network).
Finally, we notice that our choice of keeping equal the $\sigma$ values is better suited to disentagle the contribution of the competition among the media to opinion fragmentation. Varying the tolerance parameters, other routes could be possible; as an example, a very small tolerance $\sigma_{mm}$ among the media (conservative media) could keep opinions fragmented even for tolerances $\sigma_{gg}$ among the gossipers beyond the consensus transition. 

The next envisioned step will be to fine tune our model with real data from social networks platforms where main stream media directly interact with users and together change continuously the opinions' as well as the information space.

\section*{Author Contributions}
All the authors contributed equally to the paper.

\section*{Acknowledgements}

We thank US grant HDTRA1-11-1-0048, CNR-PNR National Project \textquotedblright{}Crisis-Lab\textquotedblright{} and EU FET project MULTIPLEX nr.317532. 
The contents of the paper do not necessarily reflect the position or the policy of funding parties. 
WQ wants to thank Alan Advantage Italy and Alan Advantage U.S. for their scientific support.

\bibliography{MAO}

\newpage

\newpage
\section*{Supporting Information}

\subsection*{Algorithm}

The gossip network $G_g=(V_g,E_g)$ is composed by $N=\vert V_g\vert$ gossipers; the edges $E_g$ represent social contacts among gossipers. 
Nodes of the gossip network interact with randomly selected a neighbor by applying the following rule:
Given a node $i$ that has selected a node $j$, having respectively opinion $x_i$ and $x_j$ (the opinion varies in a continuous space between 0.0 and 1.0)

\begin{center}
 \begin{algorithmic}[h!]
\FOR {all $i$ $\in$ $V_g$}  
\STATE select random neighbor $j$
\IF {$|x^t_i-x^t_j|<\sigma_{gg}$}
        \STATE $x_i^{t+1} \gets x_i^t + \mu_{gg}*(x^t_j-x^t_i)$
\ENDIF \ENDFOR
\end{algorithmic}
\end{center}
where $\sigma_{gg}$ is the threshold parameter (the distance between one is likely to consider the others opinions) and $\mu_{gg}$ is a convergence parameter varying in the interval $(0,1/2]$.

The media network $G_m=(V_m,E_m)$ is composed by $M=\vert V_m\vert$ media and follows a similar rule except for the tolerance control.
Each node $\alpha$ of the media network has an opinion $y_\alpha$ selects in its neighborhood  the one that has the highest number of followers (audience) and then apply the updating.
Given two media $i,j\in V_m$

\begin{center}
 \begin{algorithmic}[h!]
\FOR {all $\alpha$ $\in$ $V_m$}  
\STATE select most influential neighbor $\beta$
\IF {$\vert y^t_\beta - y^t_\alpha \vert <\sigma_{mm}$}
        \STATE $y_\alpha^{t+1} = y_\alpha^t 
        + \mu_{mm}*J_{\alpha\beta}*(y_\beta^t-y_\alpha^t)$
\ENDIF 
\IF {$y_\alpha^{t+1} < 0$} \STATE $y_\alpha^{t+1}=0$ \ENDIF
\IF {$y_\alpha^{t+1} > 1$} \STATE $y_\alpha^{t+1}=1$ \ENDIF
\ENDFOR
\end{algorithmic}
\end{center}
where $J_{\alpha\beta}=\pm 1$ represent the polarization among the media.

The set of edges $E_{gm} \subset V_g\times V_m$ represents the media followed by the gossipers; i.e. an edge $(i,\alpha)\in E_{gm}$ means that gossiper $i$ can be influenced by the media $\alpha$. The inter-network dynamics takes place by gossipers choosing at random a media among the ones they follow:
\begin{center}
 \begin{algorithmic}[h!]
\FOR {all $i$ $\in$ $V_g$}  
\STATE select random media $\alpha$
\IF {$|x^t_i-y^t_\alpha|<\sigma_{gm}$}
        \STATE $x_i^{t+1} \gets x_i^t + \mu_{gm}*(y^t_\alpha-x^t_i)$
\ENDIF \ENDFOR
\end{algorithmic}
\end{center}

Thus, each node of the gossip network talks with his friend and then gets informed by the media (by randomly selecting one node from the media network)
In such a mechanism: a) a leader-follower dynamics emerges among media the messages delivered by the media; b) the gossip network and the media network have a feedback loop.

\subsection*{Mean Field equations}
A simple analytical tool that help to understand the main features of network-based models like our coupled $BCM$ are the mean field (MF) approximations.

We will first sketch the MF solution in the general case of a $BCM$ subject to a time dependent external field (information from the media).  Let $\left\{ x_{i}\right\} $ be the $N$ opinions of the crowd and $\left\{ y_{k}\right\} $ the $M$ opinions (memes) of the media with $x_{i},y_{k}\in\left(0,1\right)$.
Let's suppose that at each step an individual $i$ can get information by another randomly chosen person $j$ with probability $\alpha_{g}$ or by a randomly chosen media $k$ with probability $\alpha_{m}$, with $\alpha_{i}+\alpha_{m}=1$. The individual then gets influenced by the other opinion with probability $c_{g}\left(x,y\right)$ by individuals and $c_{m}\left(x,y\right)$ by media:
\begin{equation}
x_{i}\to\left\{ \begin{array}{ccc}
x_{i}+\mu_{gg}\left(x_{j}-x_{i}\right) & with\,\, probability & \alpha_{g}\cdot c_{g}\left(x,y\right)\\
x_{i}+\mu_{gm}\left(h_{k}-x_{i}\right) & with\,\, probability & \alpha_{m}\cdot c_{m}\left(x,y\right)
\end{array}\right.
\end{equation}
where $c_g$ and $c_m$ are functions that measure the interaction strenghts among different opinions. Notice that the standard $BCM$ corresponds to $\alpha_{g}=1$ and $c_{g}\left(x,y\right)=1-\theta\left(\sigma_{gg}-\vert x-y\vert\right)$,
where $\theta\left(x\right)$ is the Heavyside theta-function and $\sigma_{gg}$ is the tolerance parameter of the $BCM$ model.
Let $p_{g}\left(x,t\right)$ and $p_{m}\left(x,t\right)$ be the probability distribution of individual opinions and of media opinions at time $t$. Rescaling the time by the average rate at which individuals receive informations, the evolution of the individual opinion distribution $p_{g}\left(x,t\right)$ is described by the master equation 
\begin{equation}
\partial_{t}p_{g}\left(x,t\right)=\int dy\int dz\, p_{g}\left(y,t\right){\displaystyle \sum_{s\in\left\{ g,m\right\} }}\alpha_{s}\, c_{s}\left(y,z\right)\, p_{s}\left(z,t\right)\left[\delta\left(x-\mu_{gs}z-\left(1-\mu_{gs}\right)y\right)-\delta\left(x-y\right)\right]\label{eq:MFevolvePindividuals}
\end{equation}
Notice that in the $BCM$ case with the standard choice $\mu_{gg}=1/2$,
eq.$\,$\ref{eq:MFevolvePindividuals} reduces to the MF equations 
\begin{equation}
\partial_{t}p_{g}\left(x,t\right)=\int dy\int_{\vert y-z\vert<\sigma_{gg}}dz\, p_{g}\left(y,t\right)p_{g}\left(z,t\right)\left[\delta\left(x-\frac{\left(y+z\right)}{2}\right)-\delta\left(x-y\right)\right]\label{eq:BenNaimMF}
\end{equation}
derived for the $BCM$ model by Ben-naim and coauthors \cite{Bennaim2003}.
Let's now consider for simplicity the case in which $c_{g}\left(x,y\right)=c_{m}\left(x,y\right)=1-\theta\left(\sigma-\vert x-y\vert\right)$
and the distribution of media opinions is time independent. It is
easy to check that to the field $p_{m}\left(x\right)=\sum\delta\left(x-y_{k}\right)$
with $\vert y_{k}-y_{l}\vert>\sigma\,\forall k,l$ there corresponds
the stationary solution $p_{g}\left(x\right)=p_{m}\left(x\right)$.
Notice that the MF approximation comes from disregarding joint correlations
among the opinions, i.e.
\begin{equation}
p\left(x_{i_{1}},x_{i_{2}},\ldots,x_{i_{L}}\right)={\displaystyle \prod_{k=1}^{L}p\left(x_{i_{k}}\right)}
\end{equation}

We will now consider the general case in which the dynamics of the media
is coupled to the dynamics of the opinions. Such a case can be described by adding to
eq.$\,$\ref{eq:MFevolvePindividuals} an equation for the field evolution
\begin{equation}
\partial_{t}p_{m}\left(y,t\right)=F\left[p_{g},p_{m},t\right]
\end{equation}
where in the MF approximation $F$ is a functional of the $p_{\alpha}$ only. 
For the media-media leader-follower dynamics, one has to understand
which is the meme that has the maximum number of followers. Disregarding
fluctuations, this can be calculated as 
\begin{equation}
h_{max}={\displaystyle \sup_{h}p_{m}\left(y\right)\int_{h-\sigma}^{h+\sigma}p_{g}\left(x\right)dx}\label{eq:MFleaderField}
\end{equation}
and the dynamics for the memes is
\begin{equation}
\partial_{t}p_{m}\left(y,t\right)=\int dz\, p_{m}\left(z,t\right)\, c_{m}\left(z,h_{max}\right)\left[\delta\left(y-\mu_{mm}z-\left(1-\mu_{mm}\right)h_{max}\right)-\delta\left(y-z\right)\right]\label{eq:MFevolvePmedia}
\end{equation}

Eq. \ref{eq:MFleaderField} makes the mean-field system of equation not amenable of simple solutions.

A further refinement of the MF approach would be needed in the case of competing media: in such a case, it is well known that even at the MF level more complicated techniques like the cavity method or the replica trick are needed to solve systems with competing interactions \cite{ParisiSGTB,MezardCM}.


\end{document}